\documentclass{article}
\usepackage[T1]{fontenc} 
\usepackage[utf8]{inputenc} 
\usepackage{ismir,amsmath,cite,url}
\usepackage{graphicx}
\usepackage{color}
\usepackage{enumitem}
\usepackage{subfig}
\usepackage[toc]{appendix}
\usepackage{lineno}

\title{What is missing in deep music generation? \\ A study of repetition and structure in popular music}

\threeauthors
  {Shuqi Dai * \thanks{* The first two authors have equal contribution.}} {Carnegie Mellon University \\ {\tt shuqid@cs.cmu.edu}}
  {Huiran Yu *} {Carnegie Mellon University \\ {\tt huiranyu@cs.cmu.edu}}
  {Roger B. Dannenberg} {Carnegie Mellon University \\ {\tt rbd@cs.cmu.edu}}

\def\authorname{S. Dai, H. Yu, and R. B. Dannenberg}

\begin{document}

\maketitle
\begin{abstract}
Structure is one of the most essential aspects of music, and music structure is commonly indicated through repetition. 
However, the nature of repetition and structure in music is still not well understood, especially in the context of music generation, and much remains to be explored with Music Information Retrieval (MIR) techniques.
Analyses of two popular music datasets (Chinese and American) illustrate important music construction principles: (1) structure exists at multiple hierarchical levels, (2) songs use repetition and limited vocabulary so that individual songs do not follow general statistics of song collections, (3) structure interacts with rhythm, melody, harmony and predictability, and (4) over the course of a song, repetition is not random, but follows a general trend as revealed by cross-entropy. These and other findings offer challenges as well as opportunities for deep-learning music generation and suggest new formal music criteria and evaluation methods. Music from recent music generation systems is analyzed and compared to human-composed music in our datasets, often revealing striking differences from a structural perspective.
\end{abstract}
\section{Introduction}\label{sec:introduction}
Structure is fundamental to music, as seen in the focus on form and analysis in music theory \cite{caplan2000, goetschius1904lessons, green1979form}, music segmentation \cite{Dannenberg2009AcousticS, Allegraud2019LearningSF, salamon2017deep}, structure analysis \cite{nieto2020, Hamanaka2014MusicalSA, Shibata2019StatisticalMS} and chorus detection \cite{paulus2010} in MIR research, and recently in the attention given to long-term dependencies in music sequence generation using deep learning techniques \cite{huang2018music, dai2021controllable}. As a basic indicator of music structure, repetition contains important music information. Music relies heavily on repetition to create internal references, coherence and structure.

Unfortunately, the nature of repetition and structure in music is still not well understood, and much remains to be explored with music information retrieval techniques. For example, while music theory may suggest that songs have distinctive motives, our work quantifies and generalizes this notion.
We will use ``structure'' to refer broadly to organizing principles in music, which are generally hierarchical and include sections, phrases and various kinds of patterns. A primary generator of structure is repetition, which includes not just music content within repeat signs but also approximate repetitions at different time scales. 

In music generation, many researchers rely on deep learning models to capture music structure and organizing principles implicitly from data.
However, repetition, especially long-term repetition structure, does not seem to emerge automatically in deep music generation. Deep learning is a promising direction, but such research should include evaluations where we can assess the successes and failures of new approaches. 
Moreover, some may argue that we do not need deep learning models to learn prevalent repetitions in music: we can produce repetition simply by generating phrases to the desired length and pasting them into a template. However, we will see that phrase structure, song structure, and other elements of music are intertwined, making this simple approach unable to reproduce characteristics of actual songs. Thus, we need a better understanding of repetition if we want machines to compose or even just listen to music in a more human way.

We aim to use formal models to explore music repetition and structure. By characterizing structural information in music, we can discover new principles of music organization and propose new challenges and evaluation strategies for music information retrieval and music generation. 

An essential aspect of repetition structure is hierarchy. We use objective data analysis to support the existence and significance of multiple levels of hierarchy in popular music. We also present a number of results that show strong interactions between structure and pitch, rhythm, harmony, entropy and cross-entropy. Simply stated, structure can help predict pitch (or rhythm, harmony, etc.) and pitch (or rhythm, harmony, etc.) can help predict structure. These findings, in turn, challenge and inform research on deep learning to model hierarchical music structure.

Another important effect of repetition is that song-specific vocabulary of rhythm and pitch patterns is limited relative to what would be expected from the entire dataset. This vocabulary serves both to unify multiple phrases within a song and distinguish songs from others.

The main contribution of this work is a better understanding of the nature of repetition in popular music. We will see that repetition exists in many forms and at different levels of hierarchy. We offer ways to quantify music repetition structure, especially as it relates to pitch and rhythm, often by measuring entropy or cross-entropy. We also reveal striking differences from a structural perspective through case studies on recent deep music generation models. These and other findings offer challenges as well as opportunities for deep-learning music generation and suggest new formal music criteria and evaluation methods. 

After describing related work in the next section, we discuss music repetition and structure in Section \ref{sec:repetition}, how it relates to deep music generation in Section \ref{sec:deeplearning}, and we present conclusions in Section \ref{sec:conclusion}.

\section{Related Work}
Repetition is a key element of music structure. Repetition is one of the three commonly used principles for segmenting music, along with novelty at segment boundaries and homogeneity within segments \cite{nieto2020}. People have developed a variety of segmentation and section detection methods based on repetition with acoustic features\cite{paulus2010, Dannenberg2009AcousticS}. 
Repetition becomes especially useful in segmenting symbolic music or lead sheet representations where timbre and texture may be lacking \cite{dai2020}.

Repetition also plays an important role in music expectation and prediction \cite{huron2006, narmour1992analysis}.
Studies of repetition and structure are common in Music Psychology \cite{margulis}.
For example, listening experiments with reordered Classical and Popular music have shown that listeners are rather insensitive to restructuring, but these results are subtle and somewhat ambiguous \cite{rolison2012}.
Music form and structure, including repetition, is also a major focus of Music Theory  \cite{caplan2000,julian2018,summach2011}.

There are many deep learning models for music generation \cite{huang2018music, roberts2018hierarchical, dhariwal2020jukebox, musenet, briot2019deep}, however, capturing repetition and long-term dependencies in music still remains a challenge. One mainstream approach is to model distribution of music via an intermediate representation (embedding), such as Variational Auto-Encoders (VAE)  \cite{roberts2018hierarchical, wei2022learning}, Generative Adversarial Networks (GANs) \cite{yu2017seqgan} and Contrastive Learning \cite{hadjeres2020vector, wei2022learning}. Due to their fixed-length representation and short-length output, it is difficult to exhibit long-term structure. Another popular trend is to use sequential models such as LSTMs and Transformers \cite{vaswani2017attention, huang2018music, musenet} to generate longer music sequences, but they still struggle to generate repetition and coherent structure on long-term time scales. Some recent work  introduces explicit structure planning for music generation, which shows that using structure information leads to better musicality \cite{dai2021controllable, medeot2018structurenet, collins2017computer}.

Current evaluation methods for music generation rarely consider repetition and structure. Deep music generation systems \cite{huang2019counterpoint, huang2018music} use objective metrics such as negative log-likelihood, cross-entropy and prediction accuracy to compare generated music with ground-truth human-composed music. But these metrics do not precisely correspond to human perception and are not reliable for musicality.
Another trend is using domain-knowledge \cite{chuan2018modeling} and musical features \cite{yang2020evaluation, elowsson2012algorithmic, dai2021personalized, sturm2017taking} such as pitch class, pitch intervals, and rhythm density to evaluate music statistically. However, most of them ignore even short patterns, and none evaluate music structure. In contrast, we offer quantitative and objective methods 
to evaluate music repetition and structure. 

\section{Repetition and Structure in Music}\label{sec:repetition}
We are interested in three main problems concerning repetition and structure in music: (a) How are repetition and structure organized hierarchically? (b)  How do different levels of structure interplay with other music elements? (c) How does repetition play out over time?

Unlike traditional music theory with case-by-case human analysis, we explore these problems in a data-driven approach. For training and testing, we use a Chinese pop song dataset POP909 \cite{pop909-ismir2020}, which has 909 pop song performances in MIDI, and an American pop song dataset PDSA \cite{Berger2020}, in MusicXML, which has 348 American pop songs originating from 1580 to 1924. We use only songs in 4/4 time to simplify analysis.

\subsection{Repetition and Structure Hierarchy}\label{sec:structure_level}
Music structure is hierarchical. It contains multiple levels of repetitions, ranging from low-level pitch and rhythm motives (patterns) to higher-level phrases (analogous to sentences) and sections (analogous to paragraphs). We describe these in more detail and use statistical and machine learning findings in the data to explore their significance.

\subsubsection{Phrases and Sections} 
\begin{figure}
 \centerline{
 \includegraphics[width=0.95\columnwidth]{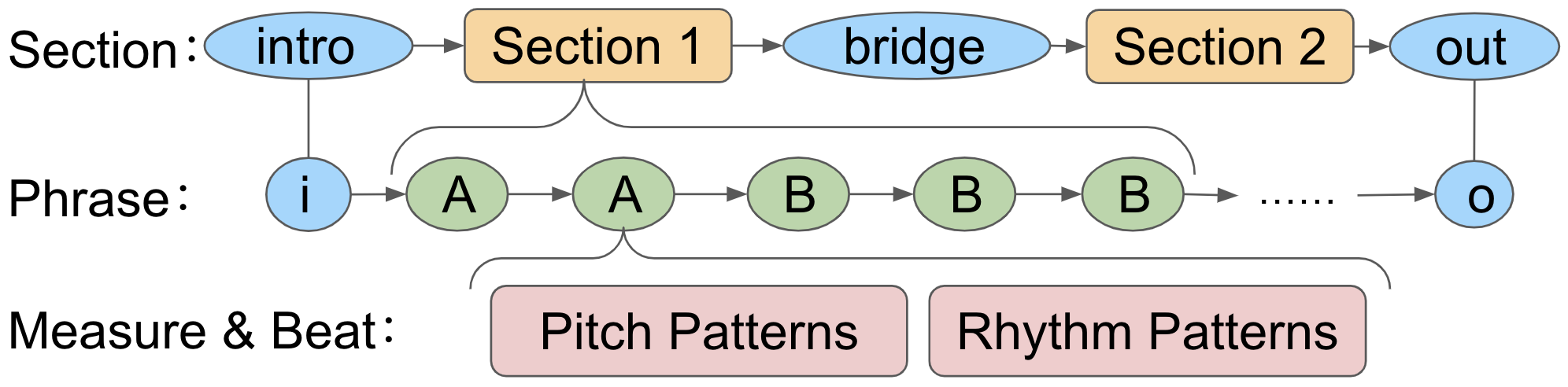}}
 \vspace{-0.5em}
 \caption{Structure hierarchy in pop music.}
 \label{fig:structure_hierarchy}
\end{figure}

Researchers\cite{dai2020} found two levels of structure in POP909: \textit{sections} and \textit{phrases}.
Phrases were identified by searching (automatically) for approximate repetitions of sequences of 4 or more measures. Non-unique phrases (those that match other phrases) often occur in sequences such as ``AABBB'' in Figure \ref{fig:structure_hierarchy} called \textit{sections}, which are by definition separated by non-repeated or non-melodic phrases. For example, Figure \ref{fig:structure_hierarchy} has an intro, two sections connected by a bridge transition, and an outro, which is a typical pop song structure. Here ``A'' can be a verse, and ``B'' can be a chorus phrase.
These phrase and section levels of structure were found to be predictive of aspects of pitch, rhythm and harmony, which shows that these two levels have significance for composition, probably for perceptual reasons. 

\begin{figure}
 \centerline{
 \includegraphics[width=0.95\columnwidth]{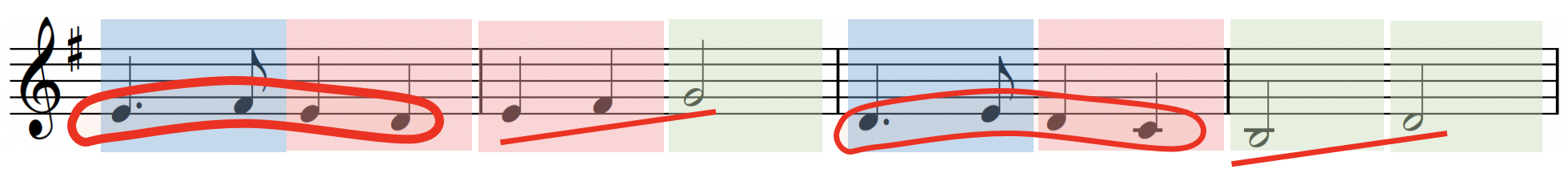}}
 \vspace{-0.5em}
 \caption{Repeated motives in a phrase in \textit{Yankee Doodle}.}
 \label{fig:motive}
 \vspace{-1em}
\end{figure}

\subsubsection{Repetition Below the Phrase Level}
\begin{figure*}
    \centering
    \includegraphics[width=\textwidth, height=4.2cm]{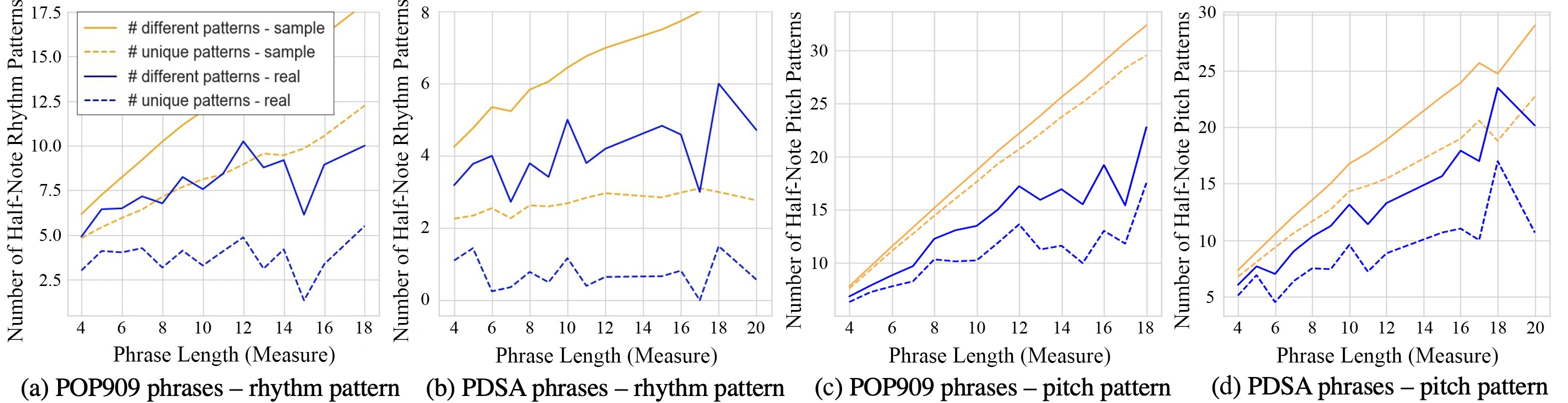}
    \vspace{-1.8em}
    \caption{Average number of different patterns (solid) and unique patterns (dashed) within a phrase vs. phrase length. (a)(b) are patterns of half-note melody note onset, (c)(d) are half-note pitch patterns. Blue lines are real phrases in the dataset. Orange lines are sampled phrases constructed by choosing each pattern at random from the entire dataset distribution. }
    \label{fig:common_pattern}
    \vspace{-0.9em}
\end{figure*}

There is at least another level of repetition below phrases (Figure \ref{fig:structure_hierarchy}). For example, in the first 8-measure phrase of the chorus in \textit{Yankee Doodle} (Figure \ref{fig:motive}), the first and second half repeat elements of rhythm and interval. The colored boxes show repeated rhythm patterns, and the red lines point out repeated pitch contours. We want to assess whether this kind of repetition is common.

We extract the phrase-level structure in PDSA using the algorithm in \cite{dai2020}, and use human-labeled structure in \cite{dai2020} for POP909. We study rhythm patterns by segmenting melody sequences into half-note onset patterns. PDSA has 54 different half-note patterns, while POP909 has all 128 possible patterns (onsets are quantized to 16th notes, and we assume an initial onset).

We claim that phrases have a limited vocabulary of onset patterns and therefore much repetition. This will of course be true necessarily if the entire dataset has a very limited vocabulary, so as a baseline for comparison, we construct random phrases by sampling half-note onset patterns from the entire dataset distribution. 
For POP909 songs, Figure \ref{fig:common_pattern}(a) shows the average number of different onset patterns in phrases of different lengths (solid lines) and also the average number of patterns that occur only once in the phrase (dashed lines). This allows us to evaluate whether patterns within phrases (blue) have similar distributions to those of the entire dataset (orange). A similar graph for PDSA songs is shown in Figure \ref{fig:common_pattern}(b).
The analysis of pitch patterns is similar. Figure \ref{fig:common_pattern}(c)(d) presents counts of melodic pitch patterns analogous to the onset pattern counts. Again, we see that real song phrases have a limited vocabulary compared to phrases assembled from random half-note units representing the entire dataset, and fewer real phrases go unrepeated.

\begin{figure}[tb]
 \centerline{
 \includegraphics[width=0.8\columnwidth, height=3cm]{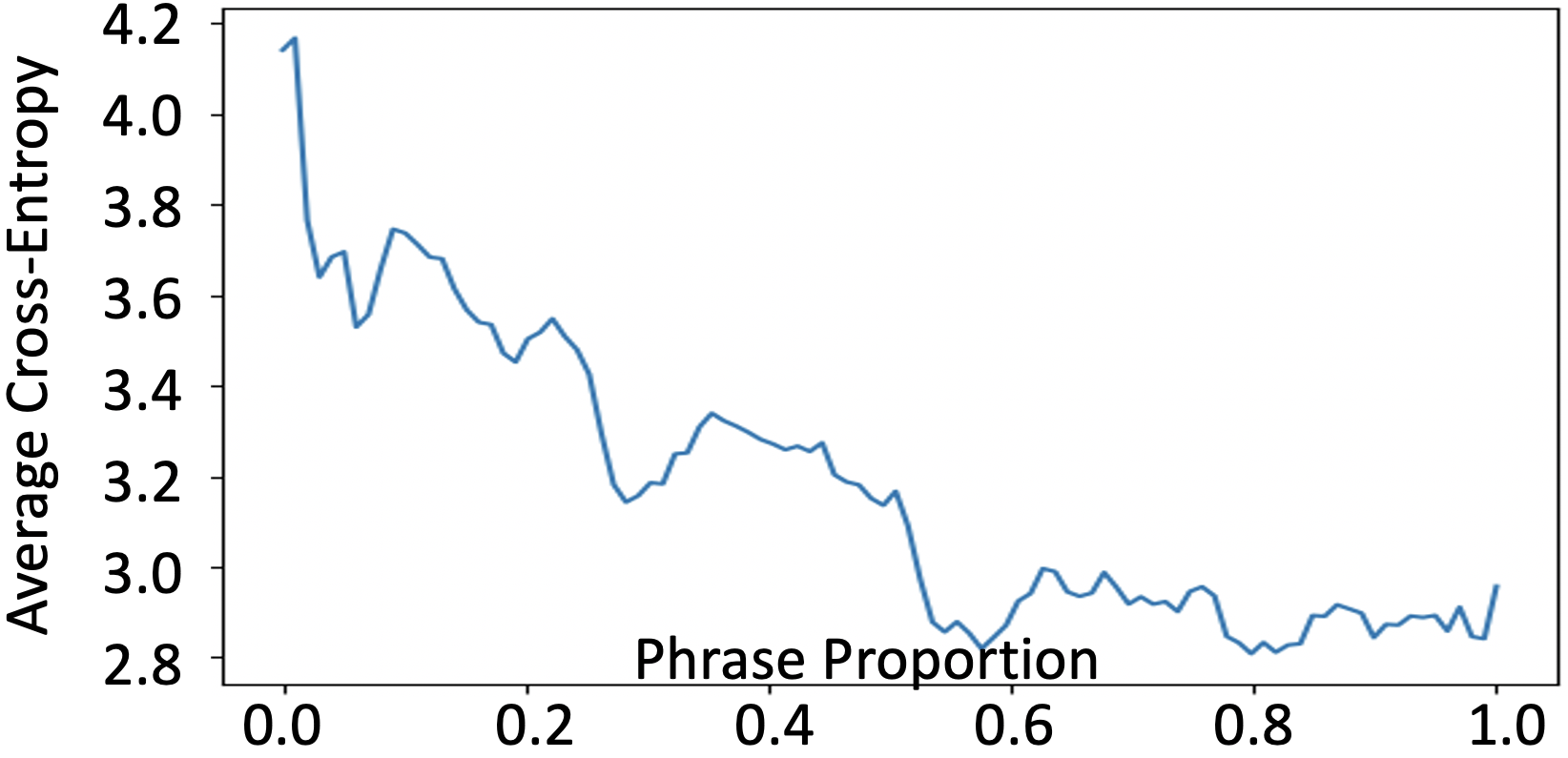}}
 \vspace{-0.5em}
 \caption{Average cross-entropy of diatonic pitch predictions over time within a phrase in POP909, using variable-order Markov models.}
 \label{fig:entropy_within_phrase}
 \vspace{-1.7em}
\end{figure}

If repetition of small patterns is frequent, variable-order Markov models \cite{begleiter2004prediction,cleary1984data} can make good song-specific online predictions by updating the model at each element of the observed sequence. Conceptually, a variable-order Markov model estimates separate Markov chains of order 0 to N based on observed state (e.g., pitch) transitions. When data is too sparse to use a higher order, the model falls back (iteratively) to the next lower order. Repetitions of different lengths are readily learned, thus prediction accuracy indicates repetition significance.
Figure \ref{fig:entropy_within_phrase} shows the average cross-entropy for pitch prediction over the length of all phrases, with phrase length normalized to 1. The clear downward trend shows the extent to which internal repetition enables better prediction.

In summary, we find abundant evidence for repetition within phrases:
\vspace{0.2em}
\begin{itemize}[leftmargin=*,noitemsep, nolistsep]
    \item Compared to sampling onset patterns at random from the POP909 distribution, real phrases have fewer distinct onset patterns, and more onset patterns are repeated in the phrase (Fig \ref{fig:common_pattern}(a)). Same trend occurs in PDSA (Fig \ref{fig:common_pattern}(b)), whose phrases contain even fewer distinct onset patterns.
    \item Rhythm patterns show a clear repetition structure within phrases. Considering each measure as a rhythm pattern, in PDSA,  7\% of phrases have the same rhythm pattern in every measure, 28\% have a repetition structure in one of the forms: ``abab,'' ``aabbaabb,'' ``aba,'' or ``abababa.''
    \item The vocabulary of pitch patterns within a song or phrase is also very limited compared to the whole dataset, implying pitch sequence repetitions within the phrase level.
    \item The cross-entropy of predicting pitches decreases over time in a phrase, suggesting within-phrase repetition.
\end{itemize}
\vspace{0.2em}
   
\subsubsection{Song-Specific Vocabulary and Common Patterns}
Motives and patterns below the phrase level create a song-specific vocabulary and have a long-term dependency throughout the whole piece. For example, in Beethoven's Symphony No.5, the first movement begins with a distinctive four-note ``short-short-short-long'' 
(the famous ``Fate Motive''). It repeats everywhere throughout the piece, and Beethoven arranged it logically to unify the entire work.

We chose variable-order Markov models to further explore patterns because (a) it is easy to tune the weights between models trained on different data, (b) models can learn different lengths of pattern repetition.
The following results show 1) song-specific vocabulary is critical in prediction, 2) there is long-term dependency between phrases.

\begin{figure}
\centering
\subfloat[\centering Including duplicated phrases in foreground training]{\includegraphics[width=0.5\columnwidth, height=3cm]{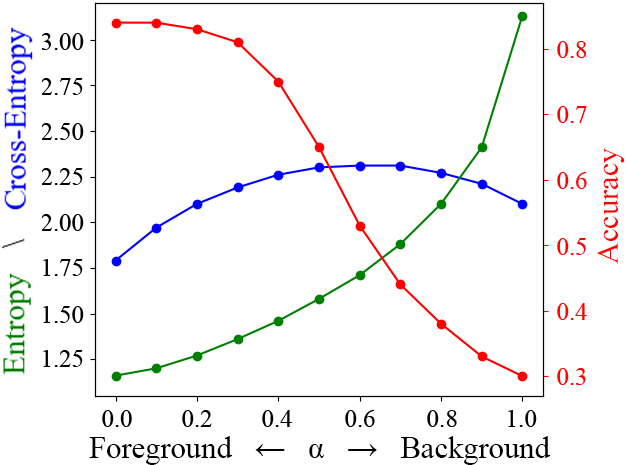}}
\subfloat[\centering Hold out repeated phrases in foreground training]{\includegraphics[width=0.5\columnwidth, height=3cm]{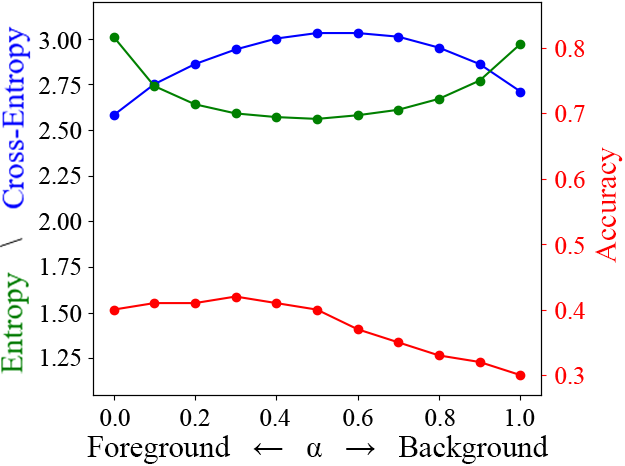}}
\vspace{-0.5em}
\caption{Average entropy, cross-entropy and accuracy of predicting the first eight diatonic pitches in each phrase, tuning between foreground and background in POP909.}
\label{fig:pop909_foreback}
 \vspace{-1em}
\end{figure}

To see the effects of song-specific vocabulary and context,  we compare the results of variable-order Markov models predicting pitch trained on many other songs (background model), versus training on the same song after removing a short segment to be predicted (foreground model). Predictions are a linear combination of the background and foreground models.
Figure \ref{fig:pop909_foreback}(a) shows that the pure foreground model always obtains the best results. One possible reason is that phrase-level repetition in the foreground allows memorization. 
Thus, we removed all the duplicated or similar phrases in the foreground training and tested only on the first eight notes in each phrase (without updating the model) to eliminate both the effects of phrase-level repetitions and within-phrase motive repetitions. With almost no information from the repeated phrase, we might expect little information within a song to help predict pitches. However, Figure \ref{fig:pop909_foreback}(b) indicates that we obtain the best prediction accuracy when incorporating 70\% foreground and 30\% background information.  
Since duplicate phrases have been removed and the first eight notes involve little within-phrase repetition, the results show that there must be a song-specific vocabulary or context that repeats across different phrases throughout the song, but not across the database (otherwise background would work better).

\subsection{Structural Influence}\label{sec:influence}

Can repetition structure be considered orthogonal to melody, harmony and rhythm generation? 
If so, we can simply generate music note-by-note, ignoring structure, and then impose structure by repeating generated sequences.
However, if there is significant interaction between structure and other facets of music, then structure is an integral part of music analysis, music modeling, and music generation. Our findings show the latter is the case.

Previous work has found systematic interactions between repetition structure and three facets of music: melody, rhythm and harmony, with interactions at both the phrase level and the section level \cite{dai2020}.
For example, chords at the ends of phrases differ from chords in the middle of phrases. Furthermore, final chords in sections differ significantly from final chords in phrases that are not at the ends of sections.
This does not mean that ``good music'' must reflect a structural hierarchy, but at least this finding offers insight into how music generation might be improved, and it raises questions for further study.

\begin{figure}[tb]
\centering
\includegraphics[width=\columnwidth]{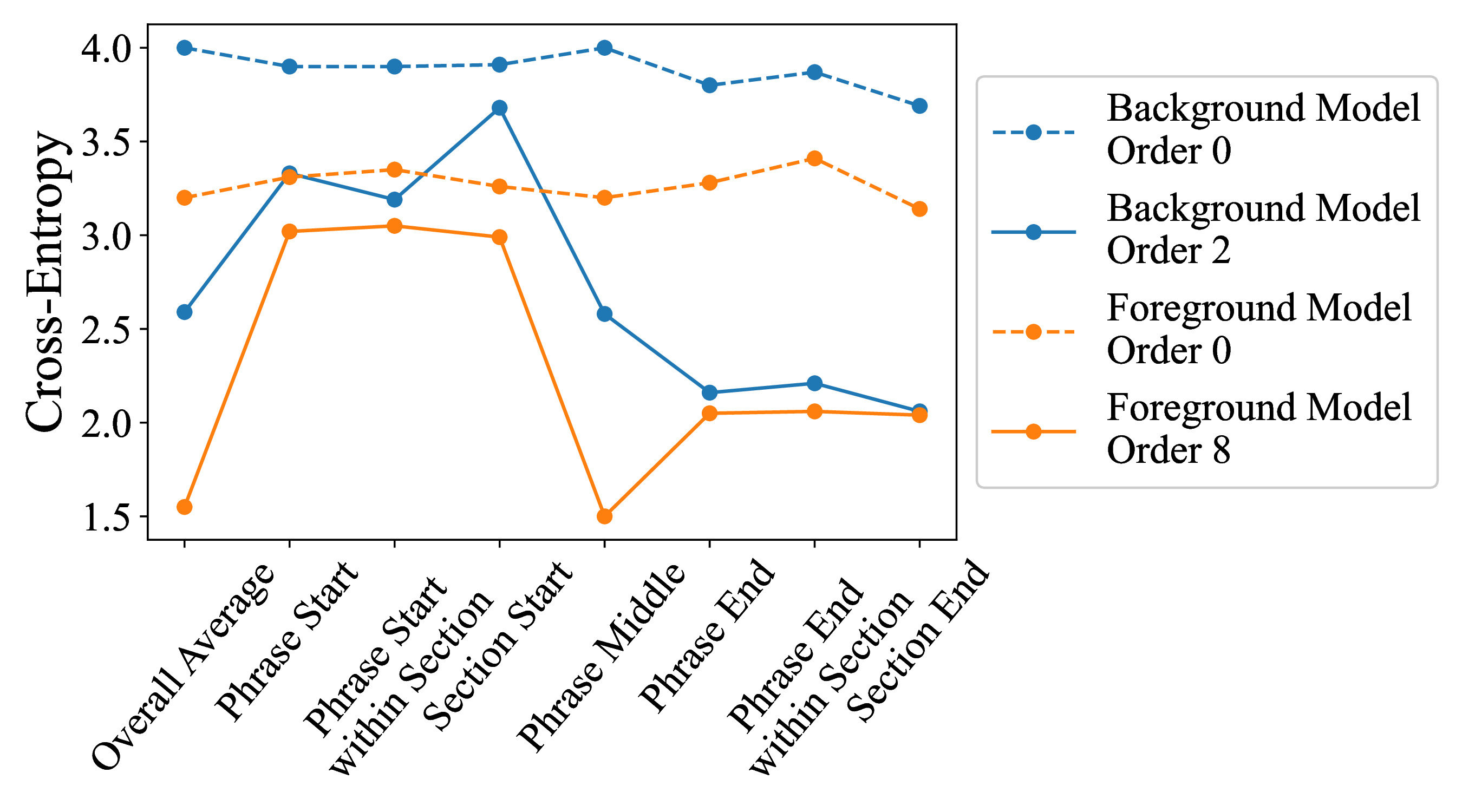}
\vspace{-2.5em}
\caption{Average cross-entropy on diatonic pitches at different structure level positions in POP909 dataset predicted by background and foreground variable Markov models.}
\label{fig:pop909_structure_pos}
 \vspace{-1.5em}
\end{figure}

In this work, we have also explored confidence and surprise in music construction to better understand the role of different levels of structure in music. In Figure \ref{fig:pop909_structure_pos}, we trained on the entire dataset (background model) of melody pitch sequences, holding out test songs; and also trained on single songs (foreground model), holding out all phrase repetitions to eliminate prediction by memorizing phrases, and holding out eight notes at a time for testing. Only order 0 (histogram) and the best-performing order maxima (2 and 8, respectively) are shown. As seen in the previous section, the foreground model predictions are better by 1 bit on average, indicating more repetition within songs (foreground) than across songs (background). 

Even though the Markov models have no positional encoding or conditions, we can still see a dramatic difference in predictability at different structure positions (Figure \ref{fig:pop909_structure_pos}). For example, using the background model, cross-entropy at the start of sections is over 3.5 (bits), while other phrase starts are about 3.2, and for most notes (phrase middle), cross-entropy is only 2.6. Using the foreground model, we see a different pattern, but predictability still varies substantially according to position in the structure.

\subsection{Repetition and Structure Over Time}
Music is not repeated randomly. After seeing different levels of hierarchy in Section \ref{sec:structure_level}, we ask: Is there a schema for repetition? How does repetition play out over time?

Huron suggests that if music is to manipulate prediction through repetition, it makes sense to repeat some of the music early on \cite{huron2006}. This affords immediate pleasure from successful prediction rather than delaying until all novel material is exhausted.
POP909 supports this hypothesis: More than 50\% of the phrases repeat immediately, and almost all phrases repeat within a quarter of the song. 

In Section \ref{sec:structure_level}, we have seen that most rhythmic and melodic patterns in a phrase are repetitions. At the song level as well, using the phrase repetition labels in POP909, we found that for 79\% of songs, 15\% to 35\% of their duration consists of new material and the rest is repetition.

\begin{figure}[tb]
 \centerline{
 \includegraphics[width=0.95\columnwidth, height=3.2cm]{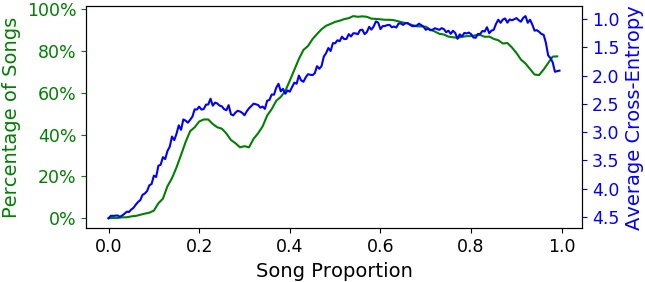}}
 \vspace{-0.8em}
 \caption{\textbf{Left} (green): Percentage of POP909 songs that have phrase repetitions at different song locations. \textbf{Right} (blue): Average prediction cross-entropy using variable-order Markov models on diatonic pitches in POP909 songs over time, and note the y-axis is inverted.}
 \label{fig:when_repeat}
  \vspace{-1.5em}
\end{figure}

Returning to our consideration of structure over time: How does surprise vary with structure? We might expect less surprise at the ends of sections to give a sense of completion or resolution. Figure \ref{fig:when_repeat} Left shows a histogram percentage of phrase-level repetition over the course of a song.
We see a relatively low repetition rate in the first 1/10 of the song. The repetition rate sharply increases as we progress to the first 1/5 of the song because, after introducing the initial materials, most songs repeat them. There is a noticeable drop around the quarter-way point where many songs introduce new material. In the second half, almost everything is a repetition or variation of what came in the first half. Finally, the graph shows novel material is often introduced near the end. In Figure \ref{fig:when_repeat} Right, we use the variable-order Markov model to calculate the average cross-entropy over time on melody pitches. To show the correspondence, we flip the vertical cross-entropy axis. Note the similarity between the trends in repetition and cross-entropy.

From these results, it is clear that repetition is not random but follows a plan in which novelty is revealed, presumably to enhance the enjoyment or effect of the music.
This is perhaps surprising because other research shows that music can be substantially rearranged without destroying positive impressions \cite{eitan2008}. Whether this organization matters to listeners or simply reflects composers' intentions requires further research.

\vspace{-0.7em}
\section{Implications for Deep Music Generation}\label{sec:deeplearning}

One application of our studies is to gain insight into
deep learning models for music generation. We can apply the analyses from Section \ref{sec:repetition} to 
melodies generated from deep learning models. Through these
case studies, we can characterize repetition and structure from deep music generation and compare to human-composed songs. To be clear, we are not claiming that any particular structure is \textit{necessary} or even \textit{good}. Our goal is to illuminate possibilities and better understand both real and generated music.
We then discuss our results and point out new directions and ideas for future work in deep music generation.

\vspace{-0.65em}
\subsection{Lack of Repetition and Structure}\label{sec:dl_casestudy}
We studied repetition and structure in deep learning generated melodies to answer three questions: 1) do melodies have multiple levels of repetition and structure? 2) do they have song-specific vocabulary and common patterns? 3) how does cross-entropy vary over time? We used two deep music generation models: One is a VAE model using representation learning \cite{wei2022learning}, chosen because it uses contrastive learning to generate longer sequences (8 bars) than other VAE models. The other model is Music Transformer \cite{huang2018music}.

\subsubsection{Do deep melodies have multiple levels of structure?}
We want to know if deep-learning-generated music has multiple levels of structure and whether any structure is reflected in pitch and rhythm. Unfortunately, most deep learning models can generate only a limited length: VAEs and GANs usually have a fixed length from 2 to 8 bars, about the length of just one phrase. Sequential models like Transformer and LSTM can generate longer sequences, but we cannot find any salient repetition comparable to repeated phrases by listening or by automated analysis. These differences from popular songs are striking.

\begin{figure}[tb]
 \centerline{
 \includegraphics[width=0.8\columnwidth]{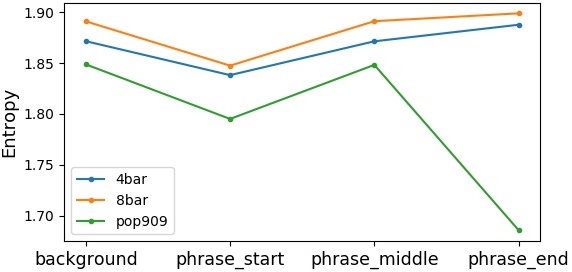}}
 \vspace{-0.5em}
 \caption{{Entropy of pitch scale degree distribution at different phrase positions, comparing generated 4-bar and 8-bar phrases from VAE with real phrases in POP909.}}
 \label{fig:VAE_pitch_structure}
  \vspace{-1.0em}
\end{figure}

Thus, the only thing we can do is to test on short generated sequences, consider them as a phrase, and see if they have phrase structure. We tested the 4 bar and 8 bar generated phrases from a VAE model \cite{wei2022learning} and used 1000 examples of each duration. Figure \ref{fig:VAE_pitch_structure} shows that there is no significant differences between the entropy of pitch scale degree distributions at the start of phrases, middle of phrases and end of phrases, but we see significant differences in the real POP909 phrases. Also, the probabilities of different pitches at different phrase-level positions only change slightly in the analyzed outputs. For example, the probability of seeing the tonic at phrase end is 35\% in real POP909 phrases, while at the other positions it is around 20\%. The probability of seeing the tonic in VAE output is in the range of 20\% to 21\% for all positions.  

\subsubsection{Do deep melodies have song-specific vocabulary and common patterns?}
Following the logic in Figure \ref{fig:common_pattern}, we count the rhythm (melody note onset) patterns for whole songs from Music Transformer and from POP909 itself. We trained the Music Transformer on 80\% of songs in POP909, using 10\% for validation and 10\% for testing. We initialized the generation with the first 2 to 6 bars of the test songs, but did not count the initialization bars as part of the generated melody. To calculate the song length, we use the total length of melodic phrases, omitting measures that are empty or have long rests.

\begin{figure}[tb]
 \centerline{
 \includegraphics[width=1.0\columnwidth]{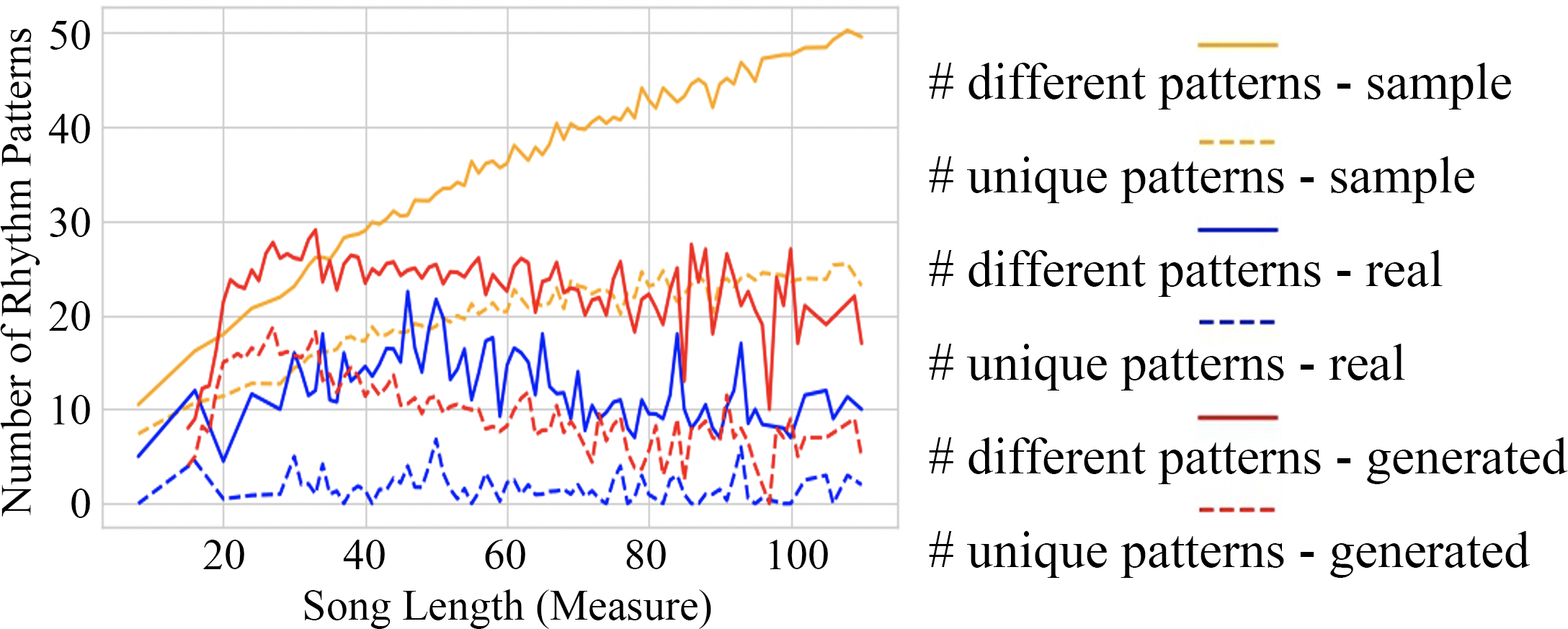}}
 \vspace{-0.5em}
 \caption{{Analysis of note onset pattern vocabulary from randomly sampled POP909 patterns (yellow), Music Transformer songs (red), and POP909 songs (blue) as a function of song length. Solid lines are the total number of different patterns, and dashed lines are the number of unique (not repeated) patterns.}}
 \label{fig:music_transformer_rhythm_vocabulary}
\end{figure}

In Figure \ref{fig:music_transformer_rhythm_vocabulary}, we see that transformer-generated results have a much higher rhythm vocabulary compared to real songs. Many people have observed that deep-learning-generated music has many outlier notes and patterns that appear once and sound like mistakes. Here we see that transformer-generated melodies have a higher number of unique rhythm patterns compared to real songs, which might explain the sense of too many outliers. 

\begin{figure}[tb]
 \centerline{
 \includegraphics[width=0.95\columnwidth]{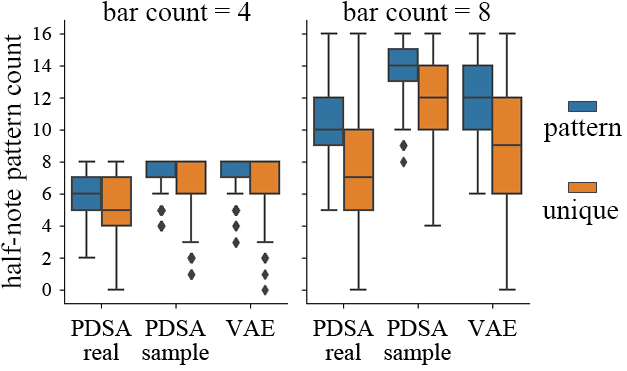}}
 \vspace{-0.7em}
 \caption{Comparison of half-note pitch pattern counts in PDSA phrases, randomly sampled PDSA pattern phrases, and phrases from VAE. The y-axis counts the number of different patterns (blue) or number of unique (not repeated) patterns (orange) found in 4 or 8 bars. The VAE generated patterns are significantly more than the patterns in the real PDSA dataset, with p-value less than $10^{-5}$.}
 \label{fig:music_vae_pitch_vocabulary}
  \vspace{-1.3em}
\end{figure}

We also analyzed half-note pitch patterns in 4-bar and 8-bar phrases from the PDSA, from random pitch patterns drawn from the entire PDSA distribution, and for the VAE model \cite{wei2022learning} (see Figure \ref{fig:music_vae_pitch_vocabulary}). Vocabulary is more limited in PDSA phrases compared to the random patterns and VAE patterns. The VAE pattern counts are closer to those of synthetic phrases and thus may lack some of the coherence, redundancy, and predictability of PDSA.

\subsubsection{Entropy over time}
\begin{figure}[tb]
 \centerline{
 \includegraphics[height=3cm]{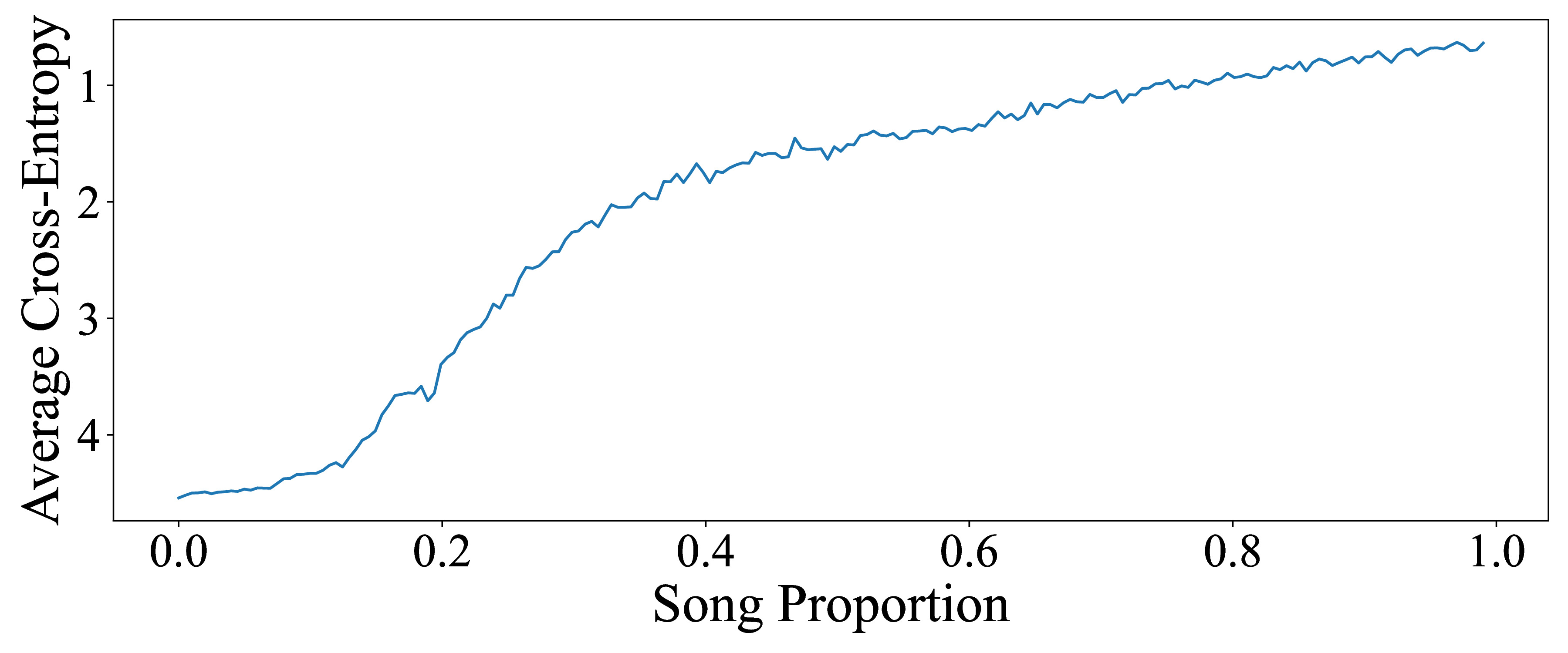}}
 \vspace{-0.5em}
 \caption{{Average prediction cross-entropy using Markov models on diatonic pitches over time in 910 generated music pieces from music transformer, y-axis inverted.}}
 \label{fig:overtime_transformer}
  \vspace{-1.5em}
\end{figure}

We also compared trends in cross-entropy of the pitch sequence over the course of songs from POP909 (Figure \ref{fig:when_repeat} blue) and Music Transformer (Figure \ref{fig:overtime_transformer}). Although both exhibit an overall cross-entropy decrease over the course of the phrase, there are important differences. POP909 cross-entropy increases at around 20\% of the song length, probably due to the introduction of novel and contrasting patterns, and also around the end.  POP909's average cross-entropy is greater than one bit per note except for a small region around 90\% of song length, while Music Transformer is below one on average for the last 30\% of the song, suggesting overly predictable sequences.

 \subsection{Discussion and New Directions}
 
Rather than learning and reproducing general statistics of datasets, we need to learn how songs strategically diverge from background or stylistic norms to create interest, surprise, and individuality. It is particularly interesting that phrases can be better predicted by relatively short phrases within the same song than by large amounts of training data from other songs. It seems plausible that songs of the same artist or same sub-genre may be more predictive than songs in general.
Our findings also reinforce previous findings that using structure in MusicFrameworks \cite{dai2021controllable} results in better human evaluations.

Examples in Section \ref{sec:dl_casestudy} suggest that we can compare generated music to real music using measures of structure, repetition and entropy. Matching these measures is not guaranteed to make music ``better,'' but we should not simply ignore clear objective differences. We would at least expect differences to be small when the task is to imitate a style or genre. We can also speculate that these measures are relevant to listener preferences even if they do not tell a complete story.

Although we have focused on popular music, repetition and hierarchical structure seem to be ubiquitous in music. Pop music, with its nearly exact repetitions, seems easier to study than Classical music where we might expect more variation, development and modulation, which make repetition less obvious. We are encouraged by our results using variable-order Markov models on pitch sequences. These studies reveal much more information than we expected and do not rely on finding wholly duplicated measures, which we used to extract phrases and sections. Perhaps this approach will be of use in Classical and other music.

\section{Conclusion}\label{sec:conclusion}

It should be no surprise that structure, repetition, pitch, rhythm, harmony and entropy are all strongly connected and interdependent. We have offered new ways to explore these connections objectively, using a data-driven approach without relying on subjective human analyses.

Among our findings are that within-song and within-phrase vocabulary and repetition are not a reflection of more general background statistics from a collection of songs. Instead, songs and phrases gain ``individuality'' through more repetition and smaller vocabulary. This has important implications for machine learning and music generation systems.

There are clear differences between measurements of real songs and those of many music generation systems, suggesting that there are important gaps to fill through new research. We hope that this work will inspire further research in the roles played by repetition and structure in music as well as methods to learn repetition and structure.

\section{Acknowledgement}
We would like to thank Ziyue Piao and Ying Yee Wong for their assistance in the early stage.

\bibliography{main}

\end{document}